\documentclass{egpubl}
\usepackage{hpg23DL}

\WsPaper           

\usepackage[T1]{fontenc}
\usepackage{dfadobe}  

\usepackage{cite}  
\BibtexOrBiblatex
\electronicVersion
\PrintedOrElectronic
\ifpdf \usepackage[pdftex]{graphicx} \pdfcompresslevel=9
\else \usepackage[dvips]{graphicx} \fi

\usepackage{egweblnk} 

\usepackage{enumitem}
\usepackage{algorithm}
\usepackage{algpseudocode}
\usepackage{amsmath}
\usepackage{amssymb}
\usepackage{gensymb}
\usepackage{tikz}
\usepackage{pgfplots}
\usepackage{contour}
\usepgfplotslibrary{patchplots}
\usetikzlibrary{shadows,backgrounds,arrows,positioning,matrix,calc,decorations.text}
\pgfplotsset{compat=newest}

\definecolor{myred}{rgb}{0.86,0.00,0.00}
\definecolor{myredlight}{rgb}{0.97,0.75,0.75}
\definecolor{myredlighter}{rgb}{0.99,0.94,0.94}
\definecolor{myredlighterr}{rgb}{1.0,0.98,0.98}
\definecolor{myblue}{rgb}{0.00,0.20,0.70}
\definecolor{mybluelight}{rgb}{0.75,0.80,0.93}
\definecolor{mybluelighter}{rgb}{0.94,0.95,0.98}
\definecolor{mybluelighterr}{rgb}{0.98,0.99,1.0}
\definecolor{mygreen}{rgb}{0.10,0.50,0.10}
\definecolor{mygreenlight}{rgb}{0.78,0.88,0.78}
\definecolor{mygreenlighter}{rgb}{0.94,0.97,0.94}
\definecolor{mygreenlighterr}{rgb}{0.99,0.99,0.99}
\definecolor{mygrey}{rgb}{0.40,0.40,0.40}
\definecolor{mygreylight}{rgb}{0.85,0.85,0.85}
\definecolor{mygreylighter}{rgb}{0.96,0.96,0.96}
\definecolor{mygreylighterr}{rgb}{0.99,0.99,0.99}
\definecolor{myorange}{rgb}{1.0,0.50,0.00}
\definecolor{myorangelight}{rgb}{1.0,0.87,0.75}
\definecolor{myorangelighter}{rgb}{1.0,0.96,0.93}
\definecolor{myorangelighterr}{rgb}{1.0,0.99,0.98}

\usepackage{listings}
\lstdefinelanguage{GLSL}%
{%
	morekeywords={%
		false,FALSE,NULL,true,TRUE,%
		__LINE__,__FILE__,__VERSION__,GL_core_profile,GL_es_profile,GL_compatibility_profile,%
		precision,highp,mediump,lowp,%
		break,case,continue,default,discard,do,else,for,if,return,switch,while,%
		void,bool,int,uint,float,double,vec2,vec3,vec4,dvec2,dvec3,dvec4,bvec2,bvec3,bvec4,ivec2,ivec3,ivec4,uvec2,uvec3,uvec4,mat2,mat3,mat4,mat2x2,mat2x3,mat2x4,mat3x2,mat3x3,mat3x4,mat4x2,mat4x3,mat4x4,dmat2,dmat3,dmat4,dmat2x2,dmat2x3,dmat2x4,dmat3x2,dmat3x3,dmat3x4,dmat4x2,dmat4x3,dmat4x4,sampler1D,sampler2D,sampler3D,image1D,image2D,image3D,samplerCube,imageCube,sampler2DRect,image2DRect,sampler1DArray,sampler2DArray,image1DArray,image2DArray,samplerBuffer,imageBuffer,sampler2DMS,image2DMS,sampler2DMSArray,image2DMSArray,samplerCubeArray,imageCubeArray,sampler1DShadow,sampler2DShadow,sampler2DRectShadow,sampler1DArrayShadow,sampler2DArrayShadow,samplerCubeShadow,samplerCubeArrayShadow,isampler1D,isampler2D,isampler3D,iimage1D,iimage2D,iimage3D,isamplerCube,iimageCube,isampler2DRect,iimage2DRect,isampler1DArray,isampler2DArray,iimage1DArray,iimage2DArray,isamplerBuffer,iimageBuffer,isampler2DMS,iimage2DMS,isampler2DMSArray,iimage2DMSArray,isamplerCubeArray,iimageCubeArray,atomic_uint,usampler1D,usampler2D,usampler3D,uimage1D,uimage2D,uimage3D,usamplerCube,uimageCube,usampler2DRect,uimage2DRect,usampler1DArray,usampler2DArray,uimage1DArray,uimage2DArray,usamplerBuffer,uimageBuffer,usampler2DMS,uimage2DMS,usampler2DMSArray,uimage2DMSArray,usamplerCubeArray,uimageCubeArray,struct,%
		gl_BackColor,gl_BackLightModelProduct,gl_BackLightProduct,gl_BackMaterial,gl_BackSecondaryColor,gl_ClipDistance,gl_ClipPlane,gl_ClipVertex,gl_Color,gl_DepthRange,gl_DepthRangeParameters,gl_EyePlaneQ,gl_EyePlaneR,gl_EyePlaneS,gl_EyePlaneT,gl_Fog,gl_FogCoord,gl_FogFragCoord,gl_FogParameters,gl_FragColor,gl_FragCoord,gl_FragData,gl_FragDepth,gl_FrontColor,gl_FrontFacing,gl_FrontLightModelProduct,gl_FrontLightProduct,gl_FrontMaterial,gl_FrontSecondaryColor,gl_InstanceID,gl_Layer,gl_LightModel,gl_LightModelParameters,gl_LightModelProducts,gl_LightProducts,gl_LightSource,gl_LightSourceParameters,gl_MaterialParameters,gl_ModelViewMatrix,gl_ModelViewMatrixInverse,gl_ModelViewMatrixInverseTranspose,gl_ModelViewMatrixTranspose,gl_ModelViewProjectionMatrix,gl_ModelViewProjectionMatrixInverse,gl_ModelViewProjectionMatrixInverseTranspose,gl_ModelViewProjectionMatrixTranspose,gl_MultiTexCoord0,gl_MultiTexCoord1,gl_MultiTexCoord2,gl_MultiTexCoord3,gl_MultiTexCoord4,gl_MultiTexCoord5,gl_MultiTexCoord6,gl_MultiTexCoord7,gl_Normal,gl_NormalMatrix,gl_NormalScale,gl_ObjectPlaneQ,gl_ObjectPlaneR,gl_ObjectPlaneS,gl_ObjectPlaneT,gl_Point,gl_PointCoord,gl_PointParameters,gl_PointSize,gl_Position,gl_PrimitiveIDIn,gl_ProjectionMatrix,gl_ProjectionMatrixInverse,gl_ProjectionMatrixInverseTranspose,gl_ProjectionMatrixTranspose,gl_SecondaryColor,gl_TexCoord,gl_TextureEnvColor,gl_TextureMatrix,gl_TextureMatrixInverse,gl_TextureMatrixInverseTranspose,gl_TextureMatrixTranspose,gl_Vertex,gl_VertexID,%
		gl_MaxClipPlanes,gl_MaxCombinedTextureImageUnits,gl_MaxDrawBuffers,gl_MaxFragmentUniformComponents,gl_MaxLights,gl_MaxTextureCoords,gl_MaxTextureImageUnits,gl_MaxTextureUnits,gl_MaxVaryingFloats,gl_MaxVertexAttribs,gl_MaxVertexTextureImageUnits,gl_MaxVertexUniformComponents,%
		abs,acos,all,any,asin,atan,ceil,clamp,cos,cross,degrees,dFdx,dFdy,distance,dot,equal,exp,exp2,faceforward,floor,fma,fract,ftransform,fwidth,greaterThan,greaterThanEqual,inversesqrt,length,lessThan,lessThanEqual,log,log2,matrixCompMult,max,min,mix,mod,noise1,noise2,noise3,noise4,normalize,not,notEqual,outerProduct,pow,radians,reflect,refract,shadow1D,shadow1DLod,shadow1DProj,shadow1DProjLod,shadow2D,shadow2DLod,shadow2DProj,shadow2DProjLod,sign,sin,smoothstep,sqrt,step,tan,texture1D,texture1DLod,texture1DProj,texture1DProjLod,texture2D,texture2DLod,texture2DProj,texture2DProjLod,texture3D,texture3DLod,texture3DProj,texture3DProjLod,textureCube,textureCubeLod,transpose,%
		rgb
	},
	sensitive=true,%
	morecomment=[s]{/*}{*/},%
	morecomment=[l]//,%
	morestring=[b]",%
	morestring=[b]',%
	moredelim=*[directive]\#,%
	moredirectives={define,defined,elif,else,if,ifdef,endif,line,error,ifndef,include,pragma,undef,warning,extension,version}%
}[keywords,comments,strings,directives]%
\definecolor{lstattrib}{rgb}{0,0.34,0}
\lstnewenvironment{cpp}[1][]{\lstset{language=c++, #1}}
    {}
\lstnewenvironment{glsl}[1][]{\lstset{language=GLSL, #1}}
    {}

\newcommand{\vndf}{D_\textrm{vis}}
\newcommand{\ndf}{D}
\newcommand{\vndfstd}{D_\textrm{vis,std}}
\newcommand{\ndfstd}{D_\textrm{std}}
\newcommand{\dd}{\textrm{d}}

\title[Sampling Visible GGX Normals with Spherical Caps]%
      {Sampling Visible GGX Normals with Spherical Caps}

\author[Jonathan Dupuy \& Anis Benyoub]
{\parbox{\textwidth}{\centering Jonathan Dupuy
        \hspace{1cm} Anis Benyoub 
        }
        \\
{\parbox{\textwidth}{\centering Intel Corporation}}
}

\begin{document}

 \teaser{
	\vspace{-0.65cm}
  	\centering
	\input{figures/teaser.tex}
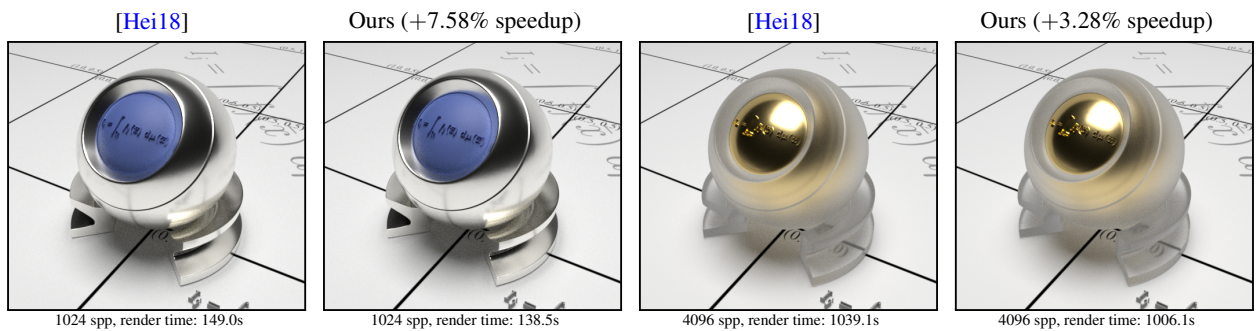
	\caption{We introduce a novel importance-sampling algorithm for GGX microfacet BSDFs. 
	Our algorithm offers the same variance reduction as the state-of-the-art method~\cite{heitz2018} 
	but with lower computational overhead. This translates into slight but systematic performance 
	gains for the rendering of, e.g., (left) rough conductors and (right) rough dielectrics. 
	Scene credit: Yasutoshi Mori for the PBRT-v4 renderer. 
	}
 	\label{fig_teaser}
}

\maketitle
\begin{abstract}
    Importance sampling the distribution of visible GGX normals requires 
    sampling those of a hemisphere. In this work, we introduce a novel method for sampling 
    such visible normals. Our method builds upon the insight that a hemispherical 
    mirror reflects parallel light rays uniformly within a solid angle shaped as a 
    spherical cap. This spherical cap has the same apex as the hemispherical mirror, 
    and its aperture given by the angle formed by the orientation of that apex 
    and the direction of incident light rays. Based on this insight, we sample GGX visible normals 
    as halfway vectors between a given incident direction and directions drawn from 
    its associated spherical cap. Our resulting implementation is even simpler than 
    that of Heitz and leads to systematic speed-ups in our benchmarks. 
   \\


\end{abstract}  
\section{Introduction}

The GGX microfacet BRDF is the most successful reflectance model in 
modern computer graphics history: 
Over the last decade, it has been implemented in many realtime game engines including 
(but not restricted to) Frostbite~\cite{hill2014}, Unreal~\cite{hill2013}, and 
Unity~\cite{lagarde2018}, as well as in offline path-tracers such as those of Walt Disney 
Animation Studios~\cite{hill2012}, Pixar~\cite{hill2013}, DreamWorks Animation~\cite{hill2017}, 
and Sony Pictures Imageworks~\cite{hill2017}. 

In this work, we introduce a novel importance sampling algorithm for the GGX microfacet BRDF. 
Our algorithm offers the same variance reduction as state-of-the-art methods albeit at faster 
execution times as shown in Figure~\ref{fig_teaser}. 
Furthermore, it is straightforward 
to integrate within an existing implementation. It can thus immediately benefit to a large 
number of physically based renderers including those just mentioned.

Our algorithm effectively offers an alternative way to sample the \emph{GGX distribution 
of visible normals (VNDF)} as described by Heitz~\cite{heitz2018}---this particular method 
lies foundation for our own algorithm and so we explain it in detail in Section~\ref{sec_previous}. 
We arrived at our alternative method by linking visible GGX normals to directions enclosed 
within solid angles shaped as spherical caps. We describe this link and our contributions throughout 
the remainder of this article as follows:

\begin{itemize}
	\item In Section~\ref{sec_model}, we link the GGX VNDF to spherical 
	caps and leverage this link to derive our novel importance sampling algorithm.
	\item In Section~\ref{sec_validation}, we validate and benchmark our algorithm against 
	that of Heitz both on the CPU and the GPU.
\end{itemize}


\begin{figure*}[h!]
	\centering
        \input{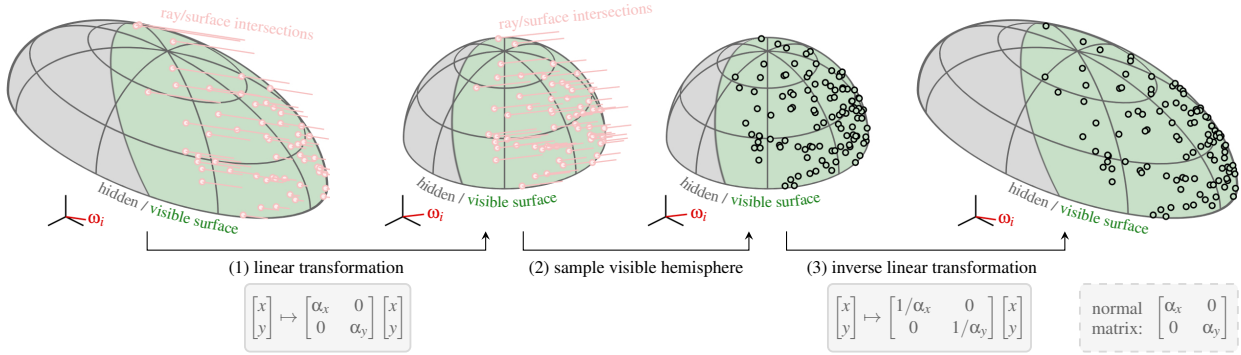}
		\vspace{-0.5cm}
	\caption{GGX VNDF sampling overview. }
	\label{fig_invariance}
\end{figure*}

\begin{figure*}[h!]
	\centering
        \input{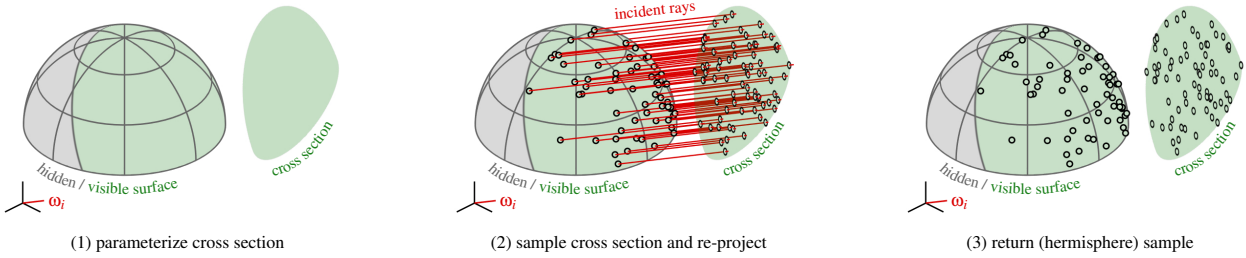}
	\caption{Sampling the visible hemisphere using its cross section.}
	\label{fig_heitz}
\end{figure*}

\section{Background on Sampling GGX Visible Normals}
\label{sec_previous}

In this section, we provide the preliminary background suitable for 
deriving a state-of-the-art GGX VNDF sampler. Note that we refrain from 
providing more general background on the GGX microfacet BRDF 
here as this is largely unnessecary for our derivations. We refer the interested 
reader to Appendix~\ref{sec_appendix_microfacet} for such supplementary information.

\paragraph*{Definition}
The GGX VNDF refers to the distribution of normals that lie at the surface 
formed by the intersection of parallel rays and a truncated ellipsoid~\cite{heitz2018}. 
This intersected surface is shaded in green in Figure~\ref{fig_invariance} and 
varies according to the direction of the parallel rays as well as the shape of 
the ellipsoid. More formally, the VNDF is 
parameterized by an incident direction $\omega_i \in \mathcal S ^2$ that,
by convention, points towards the opposite direction of the incident rays,
as well as two scaling parameters $\alpha_x > 0$ and $\alpha_y > 0$ such that 
the matrices
\begin{equation*}
    M = 
    \begin{bmatrix}
    \frac{1}{\alpha_x} & 0 & 0 \\
    0 & \frac{1}{\alpha_y} & 0 \\
    0 & 0 & 1 
    \end{bmatrix}, 
    \quad
    \textrm{and}
    \quad
    M^{-1} = 
    \begin{bmatrix}
    \alpha_x & 0 & 0 \\
    0 & \alpha_y & 0 \\
    0 & 0 & 1 
    \end{bmatrix}, 
\end{equation*}
respectively map a unit hemisphere to the ellipsoid and vice versa.
Note that this parameterization implicitely assumes that the ellipsoid 
is formed by stretching the unit hemisphere centered at the origin and lying on 
the $z=0$ plane. 
Next, we describe how to sample this distribution.

\paragraph*{Importance Sampling}
Sampling the GGX VNDF according to the method of Heitz~\cite{heitz2018} refers 
to mapping two uniform random numbers \mbox{$u_1, \, u_2 \in [0, 1)$} to a 
normal \mbox{$\omega_m$}. Intuitively, this mapping selects a ray that 
intersects the truncated ellipsoid and returns the normal lying at 
the intersection point. In practise, the GGX VNDF sampler leverages two key 
components. First, an invariance to linear transformations, which allows to 
systematically remap the intersection configuration to that of a hemisphere. 
Second, a routine to sample the visible normals of a hemisphere. 
The sampling algorithm then consists of the three following steps, which are illustrated in 
Figure~\ref{fig_invariance} (see also Listing~\ref{glsl_vndf} for a GLSL implementation):
\begin{enumerate}[label=(\arabic*)]
\item Warp the entire space using the inverse matrix $M^{-1}$. 
This maps the ellipsoid to the unit hemisphere and the incident direction as follows
\begin{align*}
    \omega_i & \mapsto \frac{M^{-1} \, \omega_i}{\|M^{-1} \, \omega_i\|}.
\end{align*}
\item Intersect the unit hemisphere based on the new incident direction and record 
the position of the intersection points. The normal associated to each intersection 
point is equal to its position. In order to sample such intersection points, Heitz 
derives an algorithm based on the parameterization of the cross section of the 
hemisphere. We illustrate this algorithm in Figure~\ref{fig_heitz} and provide a 
GLSL implementation in Listing~\ref{glsl_heitz}.
Note that this particular algorithm is the one we improve in this work.
\item Transform each intersection point back to the ellipsoid configuration using the matrix 
$M$. The normals must be warped by the inverse transpose $M^{-T}$ and effectively sample 
the GGX VNDF. 
\end{enumerate}

\begin{glsl}[caption=General GGX VNDF sampling routine., label=glsl_vndf]
vec3 SampleVndf_GGX(vec2 u, vec3 wi, vec2 alpha)
{
    // warp to the hemisphere configuration
    vec3 wiStd = normalize(vec3(wi.xy * alpha, wi.z));
    // sample the hemisphere (see implementation /*!\ref{glsl_heitz}!*/ or /*!\ref{glsl_ours}!*/)
    vec3 wmStd = SampleVndf_Hemisphere(u, wiStd);
    // warp back to the ellipsoid configuration
    vec3 wm = normalize(vec3(wmStd.xy * alpha, wmStd.z));
    // return final normal
    return wm;
}
\end{glsl}

\begin{glsl}[caption=Sampling routine for the hemisphere.
    This is a refactored version of the code distributed by Heitz~\cite{heitz2018}., label=glsl_heitz]
// Sampling the visible hemisphere using its cross section
vec3 SampleVndf_Hemisphere(vec2 u, vec3 wi)
{
    // orthonormal basis (with special case if cross product is 0)
    float tmp = wi.x * wi.x + wi.y * wi.y;
    vec3 w1 = tmp > 0.0f ? vec3(-wi.y, wi.x, 0) * inversesqrt(tmp) 
                         : vec3(1, 0, 0);
    vec3 w2 = cross(wi, w1);
    // parameterization of the cross section
    float phi = 2.0f * M_PI * u.x;
    float r = sqrt(u.y);
    float t1 = r * cos(phi);
    float t2 = r * sin(phi);
    float s = (1.0f + wi.z) / 2.0f;
    t2 = (1.0f - s) * sqrt(1.0f - t1 * t1) + s * t2;
    float ti = sqrt(max(1.0f - t1 * t1 - t2 * t2, 0.0f));
    // reprojection onto hemisphere
    vec3 wm = t1 * w1 + t2 * w2 + ti * wi;
    // return hemispherical sample
    return wm;
}
\end{glsl}

\paragraph*{Probability Distribution Function}
The probability distribution function (PDF) of the GGX VNDF can be written 
as the following linearly-transformed spherical distribution
\begin{equation}
    \label{eq_vndf}
    \vndf(\omega_m, \omega_i) 
    = 
    \underbrace{\vndfstd \left(
        \frac{M^{T} \omega_m}{\|M^{T} \omega_m\|},
        \frac{M^{-1}  \omega_i}{\|M^{-1}  \omega_i\|}
    \right)}_{\textrm{hemisphere VNDF}}
    \underbrace{
        \vphantom{\left(
            \frac{M^{T} \omega_m}{\|M^{T} \omega_m\|},
            \frac{M^{-1}  \omega_i}{\|M^{-1}  \omega_i\|}
        \right)}
        \frac{|\det \, M^T|}{\|M^{T} \omega_m\|^3}
    }_{\textrm{Jacobian}}.
\end{equation}
Here, the Jacobian is that induced by the linear transformation we introduced in the previous 
paragraph~\cite{heitz2016b}, and the term $\vndfstd$ refers 
to the VNDF of the unit hemisphere. 
This term is defined as~\cite{heitz2014a,heitz2014b}
\begin{align}
    \label{eq_vndf_std}
    \vndfstd(\omega_m, \omega_i) 
    =& 
    \frac{
        \max(\omega_m \cdot \omega_i,\,0) \, \ndfstd(\omega_m)
    }{
        \sigma_\textrm{std}(\omega_i)
    },
    \\
    \label{eq_ndf_std}
    \ndfstd(\omega_m) =& 
    \begin{cases}
        \frac{1}{\pi} & \textrm {if $z_m > 0$,} \\
        0 & \textrm{otherwise,}
    \end{cases}
    \\
    \notag
    \sigma_\textrm{std}(\omega_i) :=& \int_{\mathcal S ^2} 
    \max(\omega_m \cdot \omega_i,\,0) \, \ndfstd(\omega_m) \, \dd \omega_m
    \\
    \label{eq_sigma_std}
    =& \, \frac{1 + z_i}{2},
\end{align}
where $\ndfstd$ denotes the normal distribution function (NDF) of the 
hemisphere~\cite{walter2007}, and $\sigma_\textrm{std}$ normalizes the PDF by 
measuring the cross-sectional area of the hemisphere shown in Figure~\ref{fig_heitz}.
Note that Heitz needs to compute this area for his VNDF sampling algorithm (see line~14 in 
Listing~\ref{glsl_heitz}).

\section{Sampling GGX Visible Normals with Spherical Caps}
\label{sec_model}
In this section, we introduce our alternative method to sample 
the GGX VNDF. Our method differs from that of Heitz only in the way 
we sample the visible normals of the unit hemisphere. Our resulting GLSL implementation 
is shown in Listing~\ref{glsl_ours}. As can be seen from a direct comparison 
with the implementation of Heitz in Listing~\ref{glsl_heitz}, our code is 
considerably simpler. We emphasize that this simplification is not due to 
aggressive optimizations of the original code of Heitz but to a novel insight 
that leads to an entirely different implementation. 
We share this insight and detail the implementation it leads to in the 
remainder of this section. 

\begin{glsl}[caption=Our novel sampling routine for the hemisphere., label=glsl_ours]
// Sampling the visible hemisphere as half vectors (our method)
vec3 SampleVndf_Hemisphere(vec2 u, vec3 wi)
{
	// sample a spherical cap in (-wi.z, 1]
	float phi = 2.0f * M_PI * u.x;
	float z = fma((1.0f - u.y), (1.0f + wi.z), -wi.z);
	float sinTheta = sqrt(clamp(1.0f - z * z, 0.0f, 1.0f));
	float x = sinTheta * cos(phi);
	float y = sinTheta * sin(phi);
	vec3 c = vec3(x, y, z);
	// compute halfway direction;
	vec3 h = c + wi;
	// return without normalization (as this is done later)
	return h;
}
\end{glsl}

\subsection{Key Insight: The Specular Reflections of a Hemisphere 
are Distributed as Spherical Caps}
\begin{figure*}[t]
	\centering
        \input{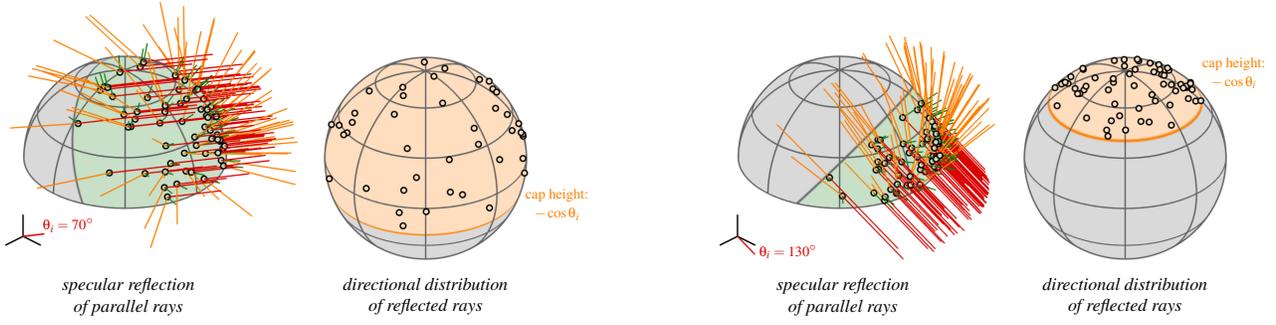}
	\caption{Main insight: a hemispherical mirror reflects parallel light rays towards directions 
	enclosed within a spherical cap.}
	\label{fig_insight}
\end{figure*}

We make the key observation that if the hemisphere used for sampling the GGX VNDF 
acted as a perfect mirror, it would reflect parallel rays uniformly within a spherical cap. 
We illustrate this property in Figure~\ref{fig_insight} for two specific 
direction of incidence---notice how the direction of incidence only 
affects the height of the cap that shapes the directional distribution 
of reflected rays. Based on this insight, we devise our sampling algorithm 
using this spherical cap rather than the cross section of the 
hemisphere as done by Heitz. Before diving into the details of our algorithm, 
we first formally prove that our insight is mathematically sound. 
This is the focus of the next subsection.

\begin{figure*}[t]
	\centering
        \input{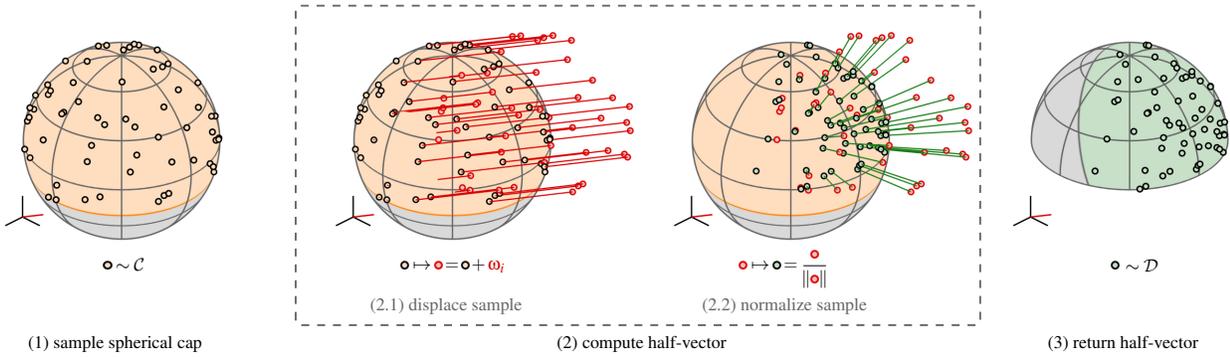}
		\vspace{-0.5cm}
	\caption{Sampling the visible hemisphere using spherical caps (our method). }
	\label{fig_cap}
\end{figure*}

\subsection{Mathematical Proof}
Let us consider that the hemisphere used for sampling the GGX VNDF 
acts as a perfect mirror, i.e., it deviates \emph{incident} 
ray directions $\omega_i$ towards \emph{outgoing} ray directions 
$\omega_o$ according to the law of specular reflection
\begin{equation}
	\label{eq_reflect}
	\omega_o = 2 (\omega_m \cdot \omega_i) \omega_m - \omega_i.
\end{equation}
In order to derive the distribution of these outgoing ray directions, we first need 
to invert Equation~\eqref{eq_reflect} to determine the location at which the 
hemisphere deflects $\omega_i$ towards $\omega_o$. This location corresponds to 
the point on the hemisphere whose normal $\omega_m$ is aligned with the half-vector 
\begin{equation}
	\label{eq_halfvector}
	\omega_h = \frac{\omega_i + \omega_o}{\|\omega_i + \omega_o\|}.
\end{equation}
Thanks to Equation~\eqref{eq_halfvector}, we can relate the density $f_p$ of outgoing 
ray directions oriented within an infinitesimal solid angle $\dd \omega_o$ to the probability 
that the incident rays will intersect the infinitesimal surface $\dd \omega_h$:
\begin{equation*}
	\notag
	f_p(\omega_o, \omega_i) \, \dd \omega_o 
	= 
	\vndfstd(\omega_h, \omega_i) \, \dd \omega_h.
\end{equation*}
Consequently, we get
\begin{align}
	\label{eq_phase_function}
	f_p(\omega_o, \omega_i) 
	&= 
	\vndfstd(\omega_h, \omega_i) 
	\left \| \frac{\dd \omega_h}{\dd \omega_o}  \right \|.
\end{align}
Now, to finalize our proof, we show that Equation~\eqref{eq_phase_function} 
reduces to a uniform density over a domain bounded by a spherical cap. We first use the 
fact that the reflection Jacobian satisfies~\cite{walter2007}
\begin{equation}
	\label{eq_jacobian}
	\left \| \frac{\dd \omega_h}{\dd \omega_o}  \right \| = 
	\frac{1}{4 \, |\omega_o \cdot \omega_h|}.
\end{equation}
Then, plugging Equations~\eqref{eq_vndf_std} and \eqref{eq_jacobian} into 
Equation~(\ref{eq_phase_function}) and using the fact that, by definition,
\mbox{$(\omega_o \cdot \omega_h) = (\omega_i \cdot \omega_h) \geq 0$}, we get
\begin{equation}
	f_p(\omega_o, \omega_i)  
	= 
	\frac{\ndfstd(\omega_h)}{4 \, \sigma_\textrm{std}(\omega_i)}.
\end{equation}
Finally, using Equations~\eqref{eq_ndf_std} and \eqref{eq_sigma_std} we 
obtain the uniform density
\begin{equation}
	\label{eq_phase_function_cap}
	f_p(\omega_o, \omega_i) = 
	\begin{cases}
		\frac{1}{2\pi \, (1 + z_i)} & \textrm{if $z_h > 0 \Rightarrow z_o > -z_i$} \\
		0 & \textrm{otherwise.}
	\end{cases}
\end{equation}
Such a uniform density is enclosed within a spherical cap 
oriented upwards and cutoff at the $z=-z_i$ plane,
thus concluding our proof.
As a final note, we mention that the normalization constant in 
Equation~\eqref{eq_phase_function_cap} corresponds to the solid angle 
of the cap that bounds the distribution.

\subsection{Sampling Algorithm}
A corollary result of our insight is that the half-vectors we retrieve from
a spherical cap cutoff at the $z = -z_i$ plane are necessarily distributed 
according to the normals of the hemisphere that are visible from direction $\omega_i$. 
This corollary result lies at the core of our sampling algorithm, which 
is illustrated in Figure~\ref{fig_cap} and consists of the following steps:
\begin{enumerate}[label=(\arabic*)]
	\item Sample a direction $\omega_o$ from a spherical cap with elevation $-z_i$.
	\item Compute the half-vector between the sample $\omega_o$ and $\omega_i$.
	\item Return this half-vector as a visible normal.
\end{enumerate}
We provide a GLSL implementation of our algorithm in Listing~\ref{glsl_ours}, which is 
meant to be used in conjunction with Listing~\ref{glsl_vndf}. An important note here 
is that we purposely omit the normalization of the half-vector at 
line~(14) of Listing~\ref{glsl_ours} (this is equivalent to avoiding step (2.2) in 
Figure~\ref{fig_cap}). We do this because 
we know that the direction is normalized later, specifically at line~(8) of 
Listing~\ref{glsl_vndf}. If one was to use Listing~\ref{glsl_ours} in a stand-alone 
fashion, then the normalization would be necessary.

\clearpage
\section{Results and Validation}
\label{sec_validation}

In this section, we validate and benchmark our method against that 
of Heitz through several experiments that we detail below.

\paragraph*{CPU Synthetic Benchmark}
We wrote a C++ program that calls Listing~\ref{glsl_vndf} 
within a loop and measured the median time it takes to run over 100 invocations. We ran 
our program on an Intel i7-13700K and measured a speed-up of 37.67\% in favor of our 
method. This result is consistent across roughness values and 
incident directions.

\paragraph*{GPU Synthetic Benchmark}
We wrote a DirectX~12 compute shader that calls Listing~\ref{glsl_vndf} 
64 times per lane and measured the median time taken for the shader to 
run over 100 invocations. 
We ran our program on an Intel~Arc~A770 and an NVIDIA~RTX~2080 and measured 
a speed-up of respectively 52.96\% and 39.25\% in favor of our method.
Again, this result is consistent across roughness values and 
incident directions.

\paragraph*{Rendering Comparisons}
We validated that our method converges to the same results as 
the method of Heitz within state-of-the-art path-tracers. 
We provide an example of such a validation in Figure~\ref{fig_teaser}, where 
we show PBRT-v4 renderings of a scene based on two different GGX microfacet BSDFs. 
As expected, our sampling algorithm does indeed converge towards the same result as Heitz.

\paragraph*{Render Profiling}
We also profiled the execution of a PBRT rendering for scenes 
of varying geometric complexity. For each scene, we set all 
materials to either rough conductors or rough dielectric. 
Under this setup we report the relative time spent for BSDF 
importance sampling in Table~\ref{tab_perf}. As demonstrated 
by the reported numbers, our sampling scheme systematically 
reduces the relative time spent for BSDF importance sampling by 
at least a factor of two compared to that of Heitz. Despite this advantage in relative time, we 
mention that, depending on the scene, one may not necessarily 
observe significant boosts in rendering times as reported, e.g., 
in Figure~\ref{fig_teaser}. 
This is because path-traced 
rendering is usually largely bottlenecked by other computations 
such as, e.g., ray intersection queries and/or memory accesses. 
Note that this is already visible in Table~\ref{tab_perf} as the relative 
time spent for BSDF sampling is systematically lower than 4\% for 
all the scenes we tested. 

\begin{table}[h]
    \begin{center}

 {
    \begin{tabular}{c||c|c||c} 
    & \multicolumn{2}{c||}{relative time} & \\ 
    scene name & \cite{heitz2018} & (ours) &  speed-up\\ 
    \hline 
    pbrt-book               & 1.34\%    & 0.54\%    & $\times 2.48$ \\
    lte-orb (conductor)     & 3.71\%    & 1.49\%    & $\times 2.49$ \\
    lte-orb (dielectric)    & 1.89\%    & 0.74\%    & $\times 2.55$ \\
    sportscar               & 0.48\%    & 0.20\%    & $\times 2.40$ \\
    \end{tabular}
}
    \end{center}
    \caption{Relative time spent for BSDF importance sampling 
    during a PBRT rendering. The scenes are available at the 
    PBRT repository: \url{https://github.com/mmp/pbrt-v4-scenes}.}
    \label{tab_perf}
\end{table}

\section{Conclusion}
We introduced a novel importance sampling algorithm for GGX microfacet BSDFs.
Our method is easy to implement and offers systematic speed-ups over that of 
Heitz.

\appendix
\section{Background on the GGX Microfacet BRDF}
\label{sec_appendix_microfacet}

The GGX microfacet BRDF refers to an analytic 
\emph{bidirectional reflectance distribution function} (BRDF), which results 
from microfacet theory. We provide here some background and the mathematical 
expressions suitable for evaluating and importance sampling such a BRDF 
via the GGX VNDF introduced in Section~\ref{sec_previous}.

\paragraph*{Microfacet Theory}
Intuitively,
microfacet theory describes surface reflectance as the result of interactions between 
incident light and microscopic mirrors (which compose the actual surface) that 
deviate this incident light. If we only consider single-scattering interactions, then 
the amount of light a surface reflects towards any given direction becomes proportional 
to the probability of intersecting a mirror suitably oriented to produce such a reflection.
This relation is formally expressed by the Cook-Torrance equation~\cite{cook1982}
\begin{equation}
    \label{eq_microfacet_theory}
    f_r(\omega_i, \omega_o) 
    = 
    \frac{
        F(\omega_h \cdot \omega_o) \, 
        G_2(\omega_h, \omega_i, \omega_o) \,
        D(\omega_h)
        }{
            4 \, \cos \theta_i \, \cos \theta_o
        },
\end{equation}
where $F$, $G_2$, and $D$ respectively denote the Fresnel term, the microfacet 
shadowing-and-masking function and the microfacet normal distribution 
function (NDF)~\cite{walter2007}. 

\paragraph*{Derivation of the GGX BRDF}
The GGX microfacet BRDF is a specialized form of Equation~\eqref{eq_microfacet_theory}, 
which follows from the two following hypothesis. The first hypothesis is that the microfacets
are oriented according to the NDF of a truncated ellipsoid~\cite{trowbridge1975,heitz2018}. 
In the spirit of 
Section~\ref{sec_previous}, we write this NDF as a function of that of a 
hemisphere. This leads to the following expression
\begin{equation}
    \ndf(\omega_m) = \underbrace{\ndfstd \left(
        \frac{M^{T} \omega_m}{\|M^{T} \omega_m\|}
    \right)}_{\substack{\textrm{hemisphere NDF} \\ \textrm{see Equation~\eqref{eq_ndf_std}}}}
    \underbrace{
        \vphantom{\left(
            \frac{M^{T} \omega_m}{\|M^{T} \omega_m\|}
        \right)}
        \frac{|\det \, M^T|}{\|M^{T} \omega_m\|^4}
    }_{\substack{\textrm{Jacobian} \\ \textrm{\cite{atanasov2022}}}},
\end{equation}
where $M$ is the matrix that maps the hemisphere to the GGX ellipsoids as 
defined in Section~\ref{sec_previous}. The second hypothesis is that the 
microfacets are subject to the Smith shadowing model~\cite{smith1967}. 
This leads to the following shadowing term~\cite{heitz2014b,dupuy2015}
\begin{align}
    G_2(\omega_m, \omega_i, \omega_o)
    &= 
    \frac{
        G_i\, G_o
    }{
        G_i + G_o - G_i\, G_o
    },
    \\
    G_k &= \, G_1 \left(\omega_m, \omega_k\right),
    \\
    \label{eq_g1}
    G_1(\omega_m, \omega_k) &= 
        \frac{
            \chi^+(\omega_m \cdot \omega_k) \, z_k
        }{
            \int_{\mathcal{S}^2} \max(\omega_m \cdot \omega_k,\,0) \, \ndf(\omega_m) \, \dd \omega_m
        },
\end{align}
where $G_1$ and $\chi^+$ respectively denote the Smith monostatic shadowing term and
the Heaviside step function. Again, we provide an analytic expression for 
Equation~\eqref{eq_g1} based on the shadowing term of the hemispere~\cite{atanasov2022}
\begin{align}
    G_1(\omega_m, \omega_i) &= G_{1, \textrm{std}} \left(
        \frac{M^{T} \omega_m}{\|M^{T} \omega_m\|},
        \frac{M^{-1}  \omega_i}{\|M^{-1}  \omega_i\|}
    \right)
    \\
    G_{1, \textrm{std}} &= 
    \frac{
        \chi^+(\omega_i \cdot \omega_m) \, z_i
    }{
        \sigma_\text{std}(\omega_i)
    },
\end{align}
where $\sigma_\text{std}$ is the cross-sectional area of the hemisphere defined in 
Equation~\eqref{eq_sigma_std} and illustrated in Figure~\ref{fig_heitz}. 

\paragraph*{Importance Sampling Using the GGX VNDF}
The GGX microfacet BRDF typically occurs in the direct illumination equation
\begin{equation}
    \label{eq_rendering}
    L(\omega_a) 
    = 
    \int_{\mathcal{S}^2} 
    L(\omega_b) 
    \, 
    f_r(\omega_b, \omega_a) 
    \, 
    |\cos \theta_b| 
    \, 
    \dd \omega_b,
\end{equation}
where $L$ denotes incident radiance. The GGX VNDF provides a way to solve 
this equation through the Monte Carlo estimator
\begin{equation}
    \label{eq_rendering_mc}
    L(\omega_a) 
    \approx
    I(\omega_a) = 
    \frac{1}{N}
    \sum_{j=0}^{N-1}
    \frac{
        L(\omega_{b_j}) 
        \, 
        f_r(\omega_{b_j}, \omega_{a}) 
        \, 
        |\cos \theta_{b_j}|
    }{
        \textrm{PDF}(\omega_{b_j}, \omega_a)
    },
\end{equation}
where $\omega_{b_j}$ denotes the $j$-th sample of the estimator. This sample is produced 
in the two following steps:
\begin{enumerate}[label=(\arabic*)]
    \item Sample the GGX VNDF from direction $\omega_a$ using Listing~\ref{glsl_vndf}. 
    This produces a normal $\omega_{m_j}$ whose PDF is given in Equation~\eqref{eq_vndf}.
    \item Use this normal to reflect the incident direction $\omega_a$ according to the 
    reflection equation $\omega_{b_j} = 2 (\omega_{m_j} \cdot \omega_a) \omega_{m_j} - \omega_a$.
\end{enumerate}
The resulting PDF is given by the VNDF weighted by the Jacobian of the reflection 
operator as provided in Equation~\eqref{eq_jacobian}:
\begin{equation}
    \textrm{PDF}(\omega_{b_j}, \omega_a) 
    = 
    \frac{
        G_1(\omega_{m_j}, \omega_a) \, D(\omega_{m_j})
    }{
        4 \, \cos \theta_a
    }.
\end{equation}
By sampling the GGX VNDF in this way, Equation~\eqref{eq_rendering_mc} effectively provides 
an estimator with state-of-the-art variance~\cite{heitz2014a}.

\bibliographystyle{eg-alpha-doi}
\bibliography{capggx}

\newcommand{\etalchar}[1]{$^{#1}$}
\begin{thebibliography}{\uppercase{AKDW22}}

\bibitem[AKDW22]{atanasov2022}
\textsc{Atanasov A., Koylazov V., Dimov R., Wilkie A.}:
\newblock Microsurface transformations.
\newblock \emph{Computer Graphics Forum 41}, 4 (2022), 105--116.
\newblock URL: \url{https://onlinelibrary.wiley.com/doi/abs/10.1111/cgf.14590},
  \href
  {http://arxiv.org/abs/https://onlinelibrary.wiley.com/doi/pdf/10.1111/cgf.14590}
  {\path{arXiv:https://onlinelibrary.wiley.com/doi/pdf/10.1111/cgf.14590}},
  \href {https://doi.org/https://doi.org/10.1111/cgf.14590}
  {\path{doi:https://doi.org/10.1111/cgf.14590}}.

\bibitem[AMG{\etalchar{*}}18]{lagarde2018}
\textsc{Abadie G., McAuley S., Golubev E., Hill S., Lagarde S.}:
\newblock Advances in real-time rendering in games.
\newblock In \emph{ACM SIGGRAPH 2018 Courses} (New York, NY, USA, 2018),
  SIGGRAPH '18, Association for Computing Machinery.
\newblock URL: \url{https://doi.org/10.1145/3214834.3264541}, \href
  {https://doi.org/10.1145/3214834.3264541}
  {\path{doi:10.1145/3214834.3264541}}.

\bibitem[CT82]{cook1982}
\textsc{Cook R.~L., Torrance K.~E.}:
\newblock A reflectance model for computer graphics.
\newblock \emph{ACM Trans. Graph. 1}, 1 (jan 1982), 7--24.
\newblock URL: \url{https://doi.org/10.1145/357290.357293}, \href
  {https://doi.org/10.1145/357290.357293} {\path{doi:10.1145/357290.357293}}.

\bibitem[DHI{\etalchar{*}}15]{dupuy2015}
\textsc{Dupuy J., Heitz E., Iehl J.-C., Poulin P., Ostromoukhov V.}:
\newblock Extracting microfacet-based brdf parameters from arbitrary materials
  with power iterations.
\newblock In \emph{Computer Graphics Forum} (2015), vol.~34, Wiley Online
  Library, pp.~21--30.

\bibitem[Hd14]{heitz2014a}
\textsc{Heitz E., d'Eon E.}:
\newblock Importance sampling microfacet-based bsdfs using the distribution of
  visible normals.
\newblock \emph{Computer Graphics Forum 33}, 4 (2014), 103--112.
\newblock URL: \url{https://onlinelibrary.wiley.com/doi/abs/10.1111/cgf.12417},
  \href
  {http://arxiv.org/abs/https://onlinelibrary.wiley.com/doi/pdf/10.1111/cgf.12417}
  {\path{arXiv:https://onlinelibrary.wiley.com/doi/pdf/10.1111/cgf.12417}},
  \href {https://doi.org/https://doi.org/10.1111/cgf.12417}
  {\path{doi:https://doi.org/10.1111/cgf.12417}}.

\bibitem[HDHN16]{heitz2016b}
\textsc{Heitz E., Dupuy J., Hill S., Neubelt D.}:
\newblock Real-time polygonal-light shading with linearly transformed cosines.
\newblock \emph{ACM Trans. Graph. 35}, 4 (jul 2016).
\newblock URL: \url{https://doi.org/10.1145/2897824.2925895}, \href
  {https://doi.org/10.1145/2897824.2925895}
  {\path{doi:10.1145/2897824.2925895}}.

\bibitem[Hei14]{heitz2014b}
\textsc{Heitz E.}:
\newblock Understanding the masking-shadowing function in microfacet-based
  brdfs.
\newblock \emph{Journal of Computer Graphics Techniques (JCGT) 3}, 2 (June
  2014), 48--107.

\bibitem[Hei18]{heitz2018}
\textsc{Heitz E.}:
\newblock Sampling the ggx distribution of visible normals.
\newblock \emph{Journal of Computer Graphics Techniques (JCGT) 7}, 4 (November
  2018), 1--13.
\newblock URL: \url{http://jcgt.org/published/0007/04/01/}.

\bibitem[HMC{\etalchar{*}}17]{hill2017}
\textsc{Hill S., McAuley S., Conty A., Drobot M., Heitz E., Hery C., Kulla C.,
  Lanz J., Ling J., Walster N., Xie F., Micciulla A., Villemin R.}:
\newblock Physically based shading in theory and practice.
\newblock In \emph{ACM SIGGRAPH 2017 Courses} (New York, NY, USA, 2017),
  SIGGRAPH '17, Association for Computing Machinery.
\newblock URL: \url{https://doi.org/10.1145/3084873.3084893}, \href
  {https://doi.org/10.1145/3084873.3084893}
  {\path{doi:10.1145/3084873.3084893}}.

\bibitem[HMD{\etalchar{*}}14]{hill2014}
\textsc{Hill S., McAuley S., Dupuy J., Gotanda Y., Heitz E., Hoffman N.,
  Lagarde S., Langlands A., Megibben I., Rayani F., de~Rousiers C.}:
\newblock Physically based shading in theory and practice.
\newblock In \emph{ACM SIGGRAPH 2014 Courses} (New York, NY, USA, 2014),
  SIGGRAPH '14, Association for Computing Machinery.
\newblock URL: \url{https://doi.org/10.1145/2614028.2615431}, \href
  {https://doi.org/10.1145/2614028.2615431}
  {\path{doi:10.1145/2614028.2615431}}.

\bibitem[MHH{\etalchar{*}}12]{hill2012}
\textsc{McAuley S., Hill S., Hoffman N., Gotanda Y., Smits B., Burley B.,
  Martinez A.}:
\newblock Practical physically-based shading in film and game production.
\newblock In \emph{SIGGRAPH 2012 Courses} (2012), ACM, pp.~10:1--7.
\newblock URL: \url{http://doi.acm.org/10.1145/2343483.2343493}, \href
  {https://doi.org/10.1145/2343483.2343493}
  {\path{doi:10.1145/2343483.2343493}}.

\bibitem[MHM{\etalchar{*}}13]{hill2013}
\textsc{McAuley S., Hill S., Martinez A., Villemin R., Pettineo M., Lazarov D.,
  Neubelt D., Karis B., Hery C., Hoffman N., Zap~Andersson H.}:
\newblock Physically based shading in theory and practice.
\newblock In \emph{SIGGRAPH 2013 Courses} (2013), ACM, pp.~22:1--8.
\newblock URL: \url{http://doi.acm.org/10.1145/2504435.2504457}, \href
  {https://doi.org/10.1145/2504435.2504457}
  {\path{doi:10.1145/2504435.2504457}}.

\bibitem[Smi67]{smith1967}
\textsc{Smith B.}:
\newblock Geometrical shadowing of a random rough surface.
\newblock \emph{IEEE Trans. on Antennas and Propagation 15}, 5 (1967),
  668--671.
\newblock \href {https://doi.org/10.1109/TAP.1967.1138991}
  {\path{doi:10.1109/TAP.1967.1138991}}.

\bibitem[TR75]{trowbridge1975}
\textsc{Trowbridge T.~S., Reitz K.~P.}:
\newblock Average irregularity representation of a rough surface for ray
  reflection.
\newblock \emph{J. Opt. Soc. Am. 65}, 5 (May 1975), 531--536.
\newblock URL:
  \url{http://www.opticsinfobase.org/abstract.cfm?URI=josa-65-5-531}, \href
  {https://doi.org/10.1364/JOSA.65.000531} {\path{doi:10.1364/JOSA.65.000531}}.

\bibitem[WMLT07]{walter2007}
\textsc{Walter B., Marschner S.~R., Li H., Torrance K.~E.}:
\newblock Microfacet models for refraction through rough surfaces.
\newblock In \emph{Proc. Eurographics Symposium on Rendering} (2007), EGSR'07,
  pp.~195--206.
\newblock URL: \url{http://dx.doi.org/10.2312/EGWR/EGSR07/195-206}, \href
  {https://doi.org/10.2312/EGWR/EGSR07/195-206}
  {\path{doi:10.2312/EGWR/EGSR07/195-206}}.

\end{thebibliography}

\end{document}